\documentclass[fleqn,usenatbib]{mnras}

\usepackage{newtxtext,newtxmath}

\usepackage{hyperref}
\usepackage[T1]{fontenc}
\usepackage{lscape}
\DeclareRobustCommand{\VAN}[3]{#2}
\let\VANthebibliography\thebibliography
\def\thebibliography{\DeclareRobustCommand{\VAN}[3]{##3}\VANthebibliography}

\usepackage{graphicx}	
\usepackage{subfig}
\usepackage{amsmath}	

\title[O($^3P$) + CO scattering]{A quantum-mechanical investigation of O($^3P$) + CO scattering cross sections at superthermal collision energies}

\author[Chhabra et al.]{
Sanchit Chhabra,$^{1}$
Marko Gacesa,$^{1,2}$\thanks{E-mail: marko.gacesa@ku.ac.ae}
Malathe S. Khalil,$^{3}$
Amal Al Ghaferi,$^{3}$
and Nayla El-Kork$^{1,2}$
\\
$^{1}$Physics Department, Khalifa University of Science and Technology, Abu Dhabi, United Arab Emirates \\
$^{2}$Space and Planetary Science Center, Khalifa University of Science and Technology, Abu Dhabi, United Arab Emirates \\
$^{3}$Mechanical Engineering Department, Khalifa University of Science and Technology, Abu Dhabi, United Arab Emirates \\
}

\date{Accepted XXX. Received YYY; in original form ZZZ}

\pubyear{2022}

\begin{document}
\label{firstpage}
\pagerange{\pageref{firstpage}--\pageref{lastpage}}
\maketitle

\begin{abstract}
The kinetics and energetic relaxation associated with collisions between fast and thermal atoms are of fundamental interest for escape and therefore also for the  evolution of the Mars atmosphere. The total and differential cross-sections of fast O($^3P$) atom collisions with CO have been calculated from quantum mechanical calculations. The cross-sections are computed at collision energies from 0.4 to 5 eV in the center-of-mass frame relevant to the planetary science and astrophysics. All the three potential energy surfaces ($^3$A', $^3$A" and 2 $^3$A" symmetry) of O($^3P$) + CO collisions separating to the atomic ground state have been included in calculations of cross-sections. The cross-sections are computed for all three isotopes of energetic O($^3P$) atoms collisions with CO. The isotope dependence of the cross-sections are compared. Our newly calculated data on the energy relaxation of O atoms and their isotopes with CO molecules will be very useful to improve the modeling of escape and energy transfer processes in the Mars' upper atmosphere.

\end{abstract}




\section{Introduction}
Carbon monoxide is the second most abundant species detected in the interstellar medium after hydrogen \citep{ehrenfreund_cami_2010}. It is found in a variety of astrophysical environments in substantial amounts, including planetary atmospheres and comets \citep{yung_demore_1999,campbell2009electron}.
For example, CO molecule plays an important role in the atmospheric convection on Jupiter \citep{doi:10.1126/science.198.4321.1031}. At Mars, CO molecule is an important atmospheric species involved in both the hydrogen and oxygen chemical cycles. It is initially produced in the Martian upper atmosphere (above 60 km) by CO$_2$ photolysis and destroyed in the lower atmosphere by hydroxyl (OH) radicals, which are the products of water vapor (H$_2$O) photolysis. Thus, CO production at Mars is balanced by its loss reaction with OH that allows it to be recycled back into CO$_2$ and act as an effective mechanism for eliminating water vapor from the atmosphere.
As a result, the abundance of CO molecules acts as a sensitive tracer of the OH-catalyzed chemistry that contributes to the stability of CO$_2$ in the martian atmosphere. Similar processes are expected to be important in balancing the atmospheres of CO$_2$-rich planets \citep{PlanetaryAtmospheresChemistryandComposition}. Similarly, on Earth, while CO is not a significant greenhouse gas, its presence in the atmosphere has a significant impact on the abundances of greenhouse gases such as CO$_2$ and CH$_4$ \citep{daniel1998climate}.



In addition to its role in the martian hydrogen and oxygen chemical cycles, molecular CO and its abundances in the martian atmosphere are important for understanding and quantifying the atmospheric escape from Mars, both at present time and over the martian history. Here, CO molecules act both as a passive participant that attenuates the escape of energetic species, as well as an important source of non-thermal C and O atoms capable of escaping from Mars to space \citep{FOX2009527}. 


At present time, the photochemical escape of atomic oxygen is a major atmospheric loss channel responsible for the most of the oxygen loss to space \citep{lillis2017photochemical,lee2020effects}. 
Strongly exothermic dissociative recombination (DR) of {O$_2$}$^{+}$ with electrons is the primary source of escaping superthermal O($^3$P), accounting for O escape \citep{FOX2009527}. It proceeds along one of the four pathways: 
\begin{align*}
                  {\mathrm{O}_2}^{+} + \mathrm{e}^{-}  \rightarrow  \mathrm{O}(^3P) +  \mathrm{O}(^3P) + 6.96 \; \mathrm{eV} \\
                                     \rightarrow  \mathrm{O}(^3P) +  \mathrm{O}(^1D) + 5.00 \; \mathrm{eV} \\
                                     \rightarrow  \mathrm{O}(^1D) +  \mathrm{O}(^1D) + 3.02 \; \mathrm{eV} \\
                                     \rightarrow  \mathrm{O}(^1D) +  \mathrm{O}(^1S) + 0.80 \; \mathrm{eV}
\end{align*}
The first two pathways of the above reaction produce superthermal O sufficiently energetic to overcome martian gravitational potential and escape to space providing they do not partially thermalize in momentum- and energy-changing collisions with other atmospheric species that they face along the way. 
Given that CO molecule is the third most abundant species in the upper atmosphere, O($^3P$)-CO collision cross sections are important parameters in evaluating the total escape rate of hot O from Mars. 


In addition to directly escaping, the superthermal oxygen atoms can also trigger the escape of other atmospheric atomic and molecular species, such as argon, helium, atomic and molecular hydrogen and its isotopes, OH, water, and CO, through collisional kinetic energy transfer \citep{https://doi.org/10.1029/2000JA000085,https://doi.org/10.1029/2009JA014055,https://doi.org/10.1029/2010GL045763,yeung_okumura_zhang_minton_paci_karton_martin_camden_schatz_2011,https://doi.org/10.1029/2012GL050904,doi:10.1063/1.4899179} or exothermal reactions with superthermal O atoms \citep{GACESA201790}. These escape processes can contribute to the supplementary non-thermal atmospheric loss. The escape rates of collisionally ejected species are estimated to be an order of magnitude smaller than that of oxygen. Nevertheless, in case of deuterium that primarily escapes through Jeans escape, the collisional ejection rate was predicted to be significant \citep{https://doi.org/10.1029/2012GL050904} and for lighter atomic and molecular trace species that do not escape significantly through other mechanisms the collisional ejection by superthermal O atoms could be the main escape channel.

Superthermal O($^3$P)+CO($^{1}\Sigma^{+}$) collisions have been studied by several authors, mostly in the context of vibrational excitations of CO. \citet{doi:10.1063/1.462105} have measured the cross sections for vibrational excitations of the ground-state CO by fast O($^3$P) at a center-of-mass collision energy of about 3.4 eV (8 km/s). By monitoring the infrared emissions of the relaxing CO they have concluded that a significant fraction of collisions produced vibrationally excited CO, with detected IR radiation from $\Delta v$ = 1 transitions of up to $v$ = 11, even though they were not able to uniquely evaluate the vibrational distribution function nor to resolve rotational emissions. Earlier, using the impulse approximation, \citet{1974PhRvA...9.1230B} predicted somewhat larger cross sections with significant excitation of higher vibrational levels for the same process. \citet{1976PhRvA..13..497K} argued that \citet{1974PhRvA...9.1230B} overestimated the excitation rates of higher vibrational levels due to the treatment of rotational excitations within the impulse approximation approach. \citet{1984cdc..rept.....R} reported the cross sections for $v$ = 1 and argued that the exchange reaction pathway could significantly contribute to higher vibrational excitations, in agreement with the conclusions of \citet{doi:10.1063/1.462105}.
\citet{doi:10.1063/1.480847} carried out a classical trajectory calculation of O($^3$P)+CO cross-sections at 3.4 eV and obtained a good agreement with the existing experiments, except at low energies \citep{1978JChPh..69.1952L,1977JChPh..66.1953K}. Their calculation was based on novel potential energy surfaces (PESs) for the three lowest energy states of the OCO complex constructed at the higher level of electron correlation than previously available effective PESs \cite{doi:10.1063/1.434206} and single-geometry calculations \citep{WILHELMWINTER1973489,doi:10.1063/1.437688,doi:10.1063/1.434270,doi:10.1063/1.434269,10.1175/1520-0469}. 
The results of \citep{doi:10.1063/1.480847} suggest ``a high degree of vibrational and rotational excitation with a nearly statistical population.''
  
More recently, \citet{brunsvold_upadhyaya_zhang_cooper_minton_braunstein_duff_2008} studied the collisional dynamics of O($^3$P) + CO collisions in a crossed molecular beams experiment supplemented by quasi-classical trajectory (QCT) calculations, again at collision energies near 3.4 eV. Their analysis suggested that, on average, about 15\% of the collision energy went into internal ro-vibrational excitations of the CO molecule as well as that the reactive collisions produced mainly forward-scattered products.

Here, we report the $^{16}$O($^3$P)+CO, $^{17}$O($^3$P)+CO, and $^{18}$O($^3$P)+CO elastic and inelastic cross sections calculated for collision energies between 0.4 eV and 5 eV, as well as differential cross sections in the entire energy range. These quantities are of interest for understanding hot oxygen transport and energy transfer at non-thermal-equilibrium conditions in planets with CO$_2$-rich atmospheres such as Mars and Venus \citep{KHARCHENKO1997107,https://doi.org/10.1029/2010JE003697}.

\section{Methodology}
\subsection{Potential Energy Surfaces}
The electronic structure of CO$_2$ has been investigated in several theoretical studies \citep{1988CPL...146..230K,1992JChPh..97.8382S,doi:10.1063/1.480847,2012JChPh.137b1101G,doi:10.1073/pnas.1213083110}.
\citet{doi:10.1063/1.480847} generated the potential energy surfaces of the three lowest electronic triplet states of CO$_2$ (one $^{3}A'$ electronic state and two $^3A''$ electronic states), which correlate to the O($^3$P) + CO($^1\Sigma^+$) asymptote, at the CASSCF-MP2 level of theory using a relatively small 631+G($d$) basis. They found that the lowest $^3A'$ and $^3A''$ states are attractive, while the second $^3A''$ state is repulsive. 
As a part of their theoretical study of CO$_2$ photolysis, \citet{doi:10.1073/pnas.1213083110} constructed the lowest 6 singlet and 6 triplet PESs, at a higher level of theory than previously available. For the lowest $^3A'$ and two lowest $^3A''$ electronic states that correlate to the ground state of CO and the ground state of O atom, \citet{schwenke2016collisional} constructed new potential energy surfaces more suitable for accurately describing the collision dynamics at superthermal energies. A number of improvements include a non-equidistant coordinate grid more suitable for the geometries relevant in dynamics, a more complete description of the O+C+O asymptotic limit, addition of scalar relativistic effects, and a number of other modifications. 
\citet{schwenke2016collisional} used their PESs in quasi-classical trajectory (QCT) nuclear dynamics calculations of thermal dissociation rate coefficients of the complex over a wide range of temperatures. 
Notably, semi-empirical variations of the surfaces were used to successfully construct accurate high-temperature line lists for CO$_2$ isotopologues \citep{doi:10.2514/6.2018-3768}. Thus, in this study, we adopt the surfaces of \cite{schwenke2016collisional}.

\subsection{Quantum dynamics}
The total cross section $\sigma_{j,j'}$, for a transition from an initial rotational state $j$ to a final rotational state $j'$, where $(j,l)$ and $(j',l')$ are the rotational and angular quantum numbers of the rigid rotor, respectively, can be expressed in terms of the scattering matrix $S^{J}_{jj';ll'}$ for the total rotational quantum number $J$ of the atom-molecule system as a sum \citep{doi:10.1098/rspa.1960.0125}
\begin{multline}{\label{eq01}}
\sigma_{j,j'} = \frac{\pi}{k_{j'j}^2 (2j+1)} \sum_{J=0}^{J_\mathrm{max}}
 \sum_{l=|J-j|}^{J+j} \sum_{l'=|J-j'|}^{J+j'} (2J+1)
 |\delta_{jj'} \delta_{ll'} - S_{jj';ll'}^{J}|^2 ,
\end{multline}
where $k^2_{j'j}$ = $2\mu/{\hbar^2} \left(E_{j} - {\hbar^2} j'(j'+1)/(2I) \right)$ is the channel wave number in the incoming channel $j$. The total energy of the system when the rotor is in the state $j$ is given by $E_j$ = $E + \hbar^2 j(j+1)/{2I}$, where $E$ is the kinetic energy of the incident particle in the center-of-mass system and $I$ is the moment of inertia of the rotor \citep{10.1093/mnras/staa1086,chhabra2019quantum,chhabra2018ab}. Note that $J$ denotes the total angular momentum of the system, $\vec{J}$ = $\vec{l}$ + $\vec{j}$, where $\vec{l}$ and $\vec{j}$ are the orbital angular momentum of the O-CO complex and rotational angular momentum of CO system, respectively. In practical applications, the sum over $J$ in Eq.(\ref{eq01}) is evaluated up to $J_\mathrm{max}$, where $J_\mathrm{max}$ is selected such that the contribution from the subsequent terms to the cross section is not significant.

A number of approximate methods for evaluating the above expression are available. The coupled state (CS) approximation \citep{doi:10.1063/1.1681388} lowers the computational time by ignoring the Coriolis coupling between different values of $\Omega$, the projection of the angular momentum quantum number of the diatom along the body fixed axis. Within the CS approximation, the total cross section is given by
\begin{multline}{\label{eq02}}
\sigma_{j,j'}(E_k) = \frac{\pi}{k_{j'j}^2(2j+1)} \sum_{J=0}^{J_\mathrm{max}}(2J+1)\\
 \times \sum_{\Omega=0}^{\Omega_\mathrm{max}}(2-\delta_{\Omega 0})
 |\delta_{jj'} - S_{jj'}^{J \Omega}(E_k)|^2
\end{multline}
where $\Omega_\mathrm{max} = 0, 1, ... , \mathrm{max}(J,j)$. 
The total cross section from a given initial state is given by
\begin{equation}{\label{eq03}}
\sigma_{j}(E_k) = \sum_{j'}\sigma_{j,j'}(E_k)
\end{equation}
where $j'$ = 0,....$j_\mathrm{max}$. The DCSs are calculated as
\begin{equation}{\label{eq04}}
Q_{jj'}(\theta,E) = d\sigma_{j,j'}(\theta,E)/d\Omega \, .
\end{equation}
The total DCSs from a given initial state are calculated as
 \begin{equation}{\label{eq05}}
Q_{j}(\theta,E) = \sum_{j'}d\sigma_{j,j'}(\theta,E)/d\Omega \, .
\end{equation}
Here, d$\Omega$ = sin$\theta$ d$\theta$ d$\phi$ is the solid angle element and $\theta$ is the scattering angle between the center-of-mass velocity vector of the initial O atom and final CO molecule defined such that $\theta$ = 0$^{\circ}$ and 180$^{\circ}$ correspond to the forward and backward scattering, respectively. 

\begin{figure}  
\begin{center}
\includegraphics [height=0.2\textwidth]{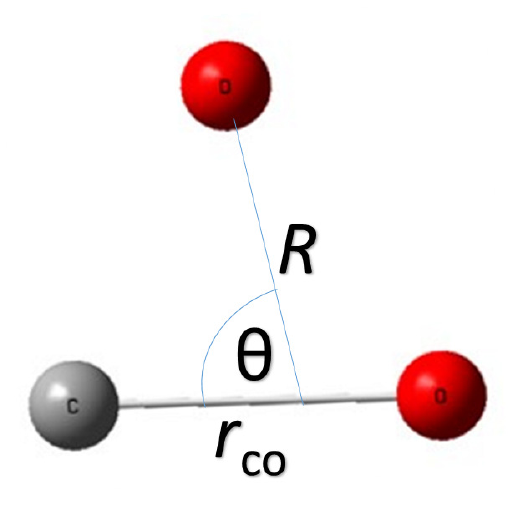}
{\caption{\label{fig:co}Coordinate system for O + CO complex.}}
\end{center}
\end{figure}

\begin{figure*}  
\begin{center}
  \subfloat{\includegraphics [height=0.40\textwidth]{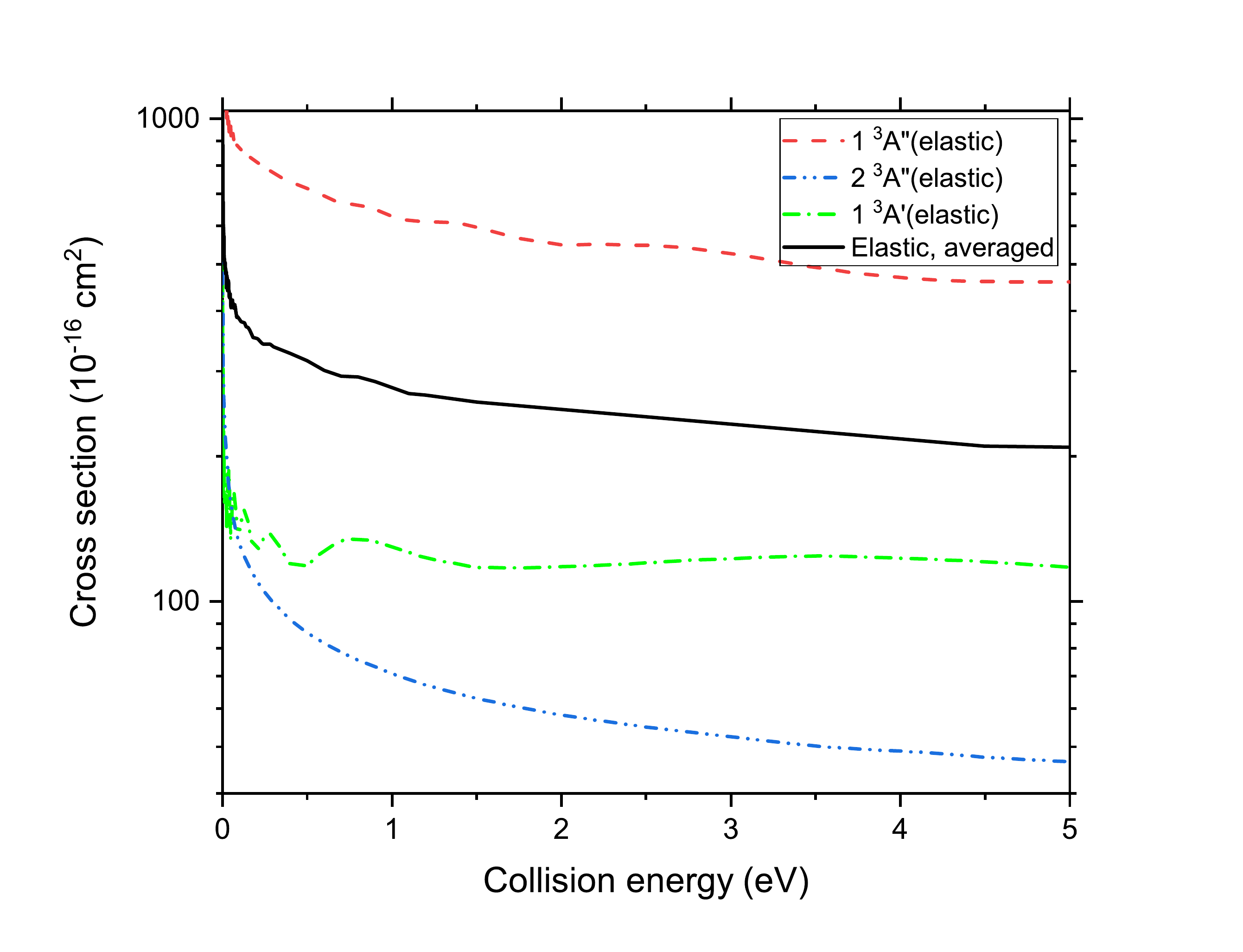}}
  \subfloat{\includegraphics 
  [height=0.40\textwidth]{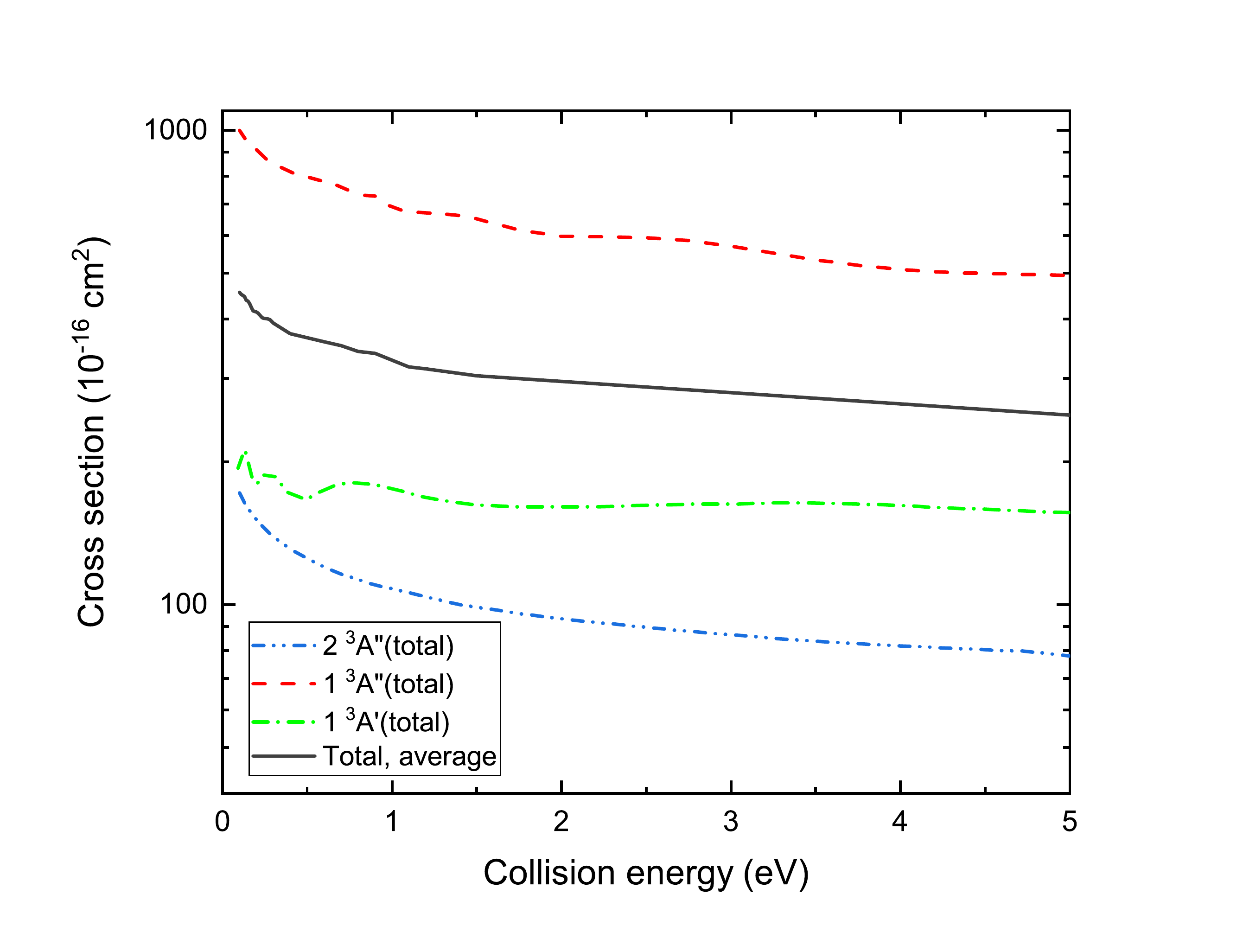}}
  {\caption{\label{fig:elastic}Elastic (left panel) and total (right panel) cross sections in dependence on the collision energy for $^{16}$O($^{3}P$) + CO scattering. The cross sections calculated for each of the three potential energy surfaces ($^3A'$, $1^3A''$, and $2^3A''$) are also shown.}}
\end{center}
\end{figure*}

\section{Results and Discussion}
\subsection{Elastic and total cross sections}
The cross sections obtained in this work are calculated using the molecular scattering code MOLSCAT \citep{hutson1994molscat}. The coupled-channel (CC) equations are integrated by adopting the modified log-derivative Airy propagator \citep{doi:10.1063/1.452154}.
The scattering S-matrices, and the resulting cross sections, are computed independently on each of the three triplet potential energy surfaces and the surface hopping is neglected. The contribution to each process was multiplied by their statistical weight due to electronic degeneracy, which amounts to the factor of 1/3: 
\begin{equation}
   \sigma = \frac{1}{3} \left[ \sigma(^{3}A') + \sigma(1^{3}A'') + \sigma(2^{3}A'') \right]
\end{equation}
In our case, this approximation is justified  because the generated $^3A'$ and $1^{3}A''$ electronic states which influence the collision dynamics below 5 eV are not permitted to interact in geometries that possess $C_S$ symmetry (where most of the collision dynamics occurs) by symmetry, nor can they interact in $C_{2v}$ geometries where they correlate to $^{3}B_2$ and $^{3}A_2$ electronic states, respectively. 
There is a possibility that these electronic states can interact in $D_{\infty h}$ geometries that exist at higher energies, above 5 eV, which falls outside of the regime that we study in this work. The $2^{3}A''$ electronic state can interact with the $1^{3}A''$ state, but only at energies well beyond threshold, and only at dissociation, where they are close to degenerate.

We carried out scattering calculations at total energies up to 5 eV on a non-equidistant grid in energy such that the grid density between the consecutive energy points is increased with the energy. Moreover, for inelastic (elastic) rotational transitions, the off-diagonal (diagonal) tolerance is set to OTOL = 0.005{\AA}$^2$ (DTOL = 0.1 {\AA}$^2$). This grants the limit value of the total angular momentum large enough for a best convergence of cross sections. The propagation was performed from $R$=0.8 {\AA} to a maximum distance of $R$=30 {\AA} .

We solved the scattering problem for low collision energy (0.4 eV) to high collision energy (5 eV).  The scattering calculations were performed using the MOLSCAT nonreactive scattering program appropriate to the current system. The CO molecule has a low rotational constant. CO molecule has large numbers of closely spaced rotational levels. In order to make the calculations affordable, an appropriate basis set is necessary to account for CO's large numbers of closely spaced rotational levels. Rotational levels $j$ = 0 $-$ 70 are included in the calculations for $v$ = 0 vibrational level. The highest rotational level has energy 9403.761 cm$^{-1}$.  We included sufficient total angular momentum partial waves $J$  in order to ensure that the cross sections converged \citep{D1CP04273D,doi:10.1063/1.5058126,kushwaha2020interaction,chhabra2020ultracold}. As collision energy increases, $J$ increases. Highest value of $J$ obtained is 1500. Parameters used in the MOLSCAT calculations are given in Table \ref{tab:table1}. State to state cross sections ($\sigma_{j=0,j'}$) for initial 30 levels as a function of selected energies for $^{16}$O with CO are given in Table \ref{tab:table5}.


Computed elastic cross sections as a function of collision energy are given in Figure \ref{fig:elastic}, as well as listed in Table \ref{tab:table3} for selected energies. Both in the case of elastic and total (elastic+inelastic) cross sections, the contribution from the $1^{3}A''$ surface is the most significant, while the $2^3A''$ surface contributes the least. 

\begin{table} 
\caption{\label{tab:table1}Parameters used in the MOLSCAT code.}
\begin{tabular}{cc}
\cline{1-2}
  $j_\mathrm{max}$ = 70                & $\mu_{16}$ = 10.18291037 a.m.u   \\
 $\mu_{17}$ = 10.5777777 a.m.u  & $\mu_{18}$ = 10.957067 a.m.u     \\
  $R_\mathrm{min}$ = 0.8 {\AA}  & $R_\mathrm{max}$ = 30 {\AA}      \\
       DTOL = 0.1 {\AA$^2$}     & OTOL = 0.005{\AA$^2$}            \\  \cline{1-2}
\end{tabular}
\end{table}

\begin{figure*} 
\begin{center}
\subfloat{\includegraphics 
[height=0.40\textwidth]{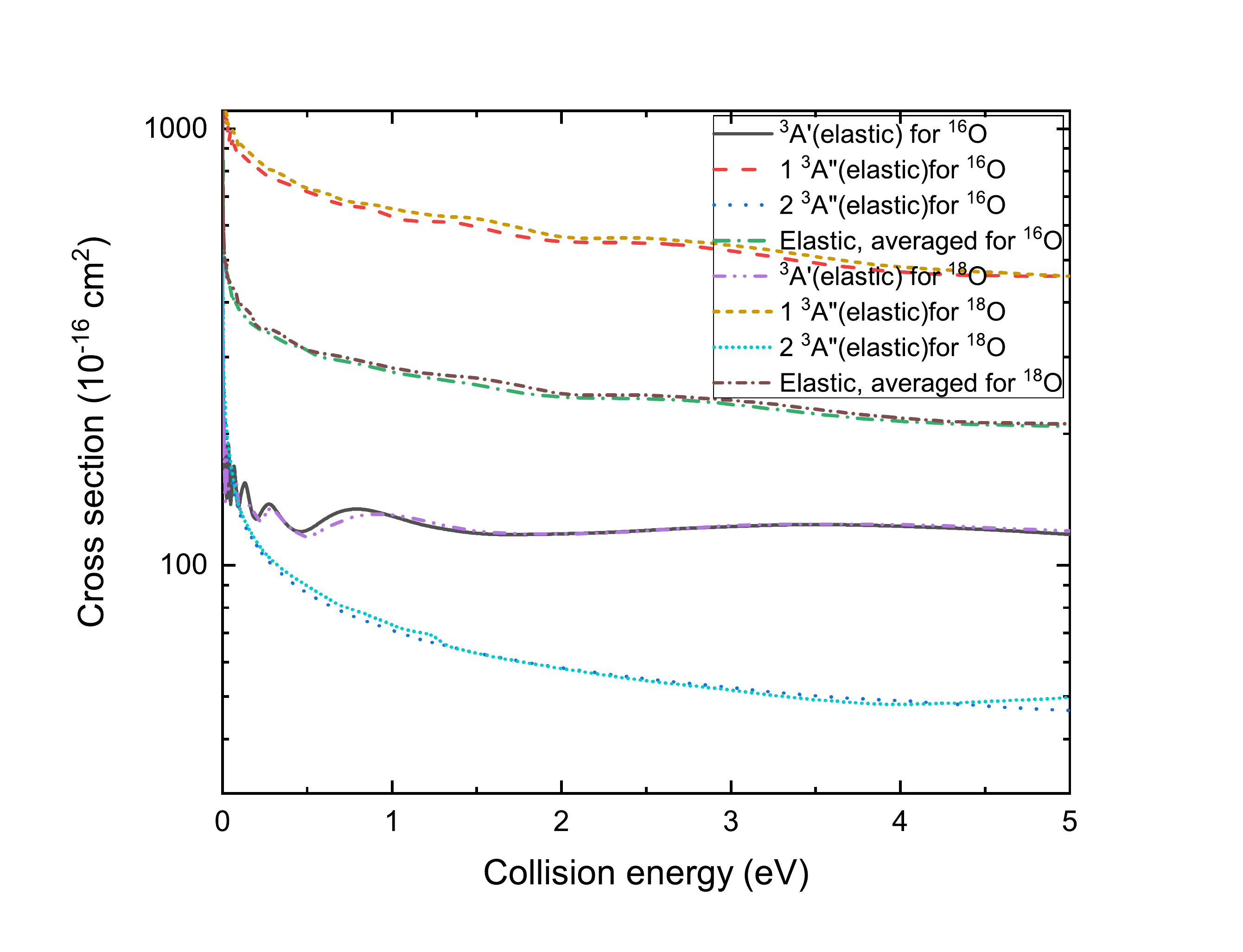}}
\subfloat{\includegraphics 
[height=0.40\textwidth]{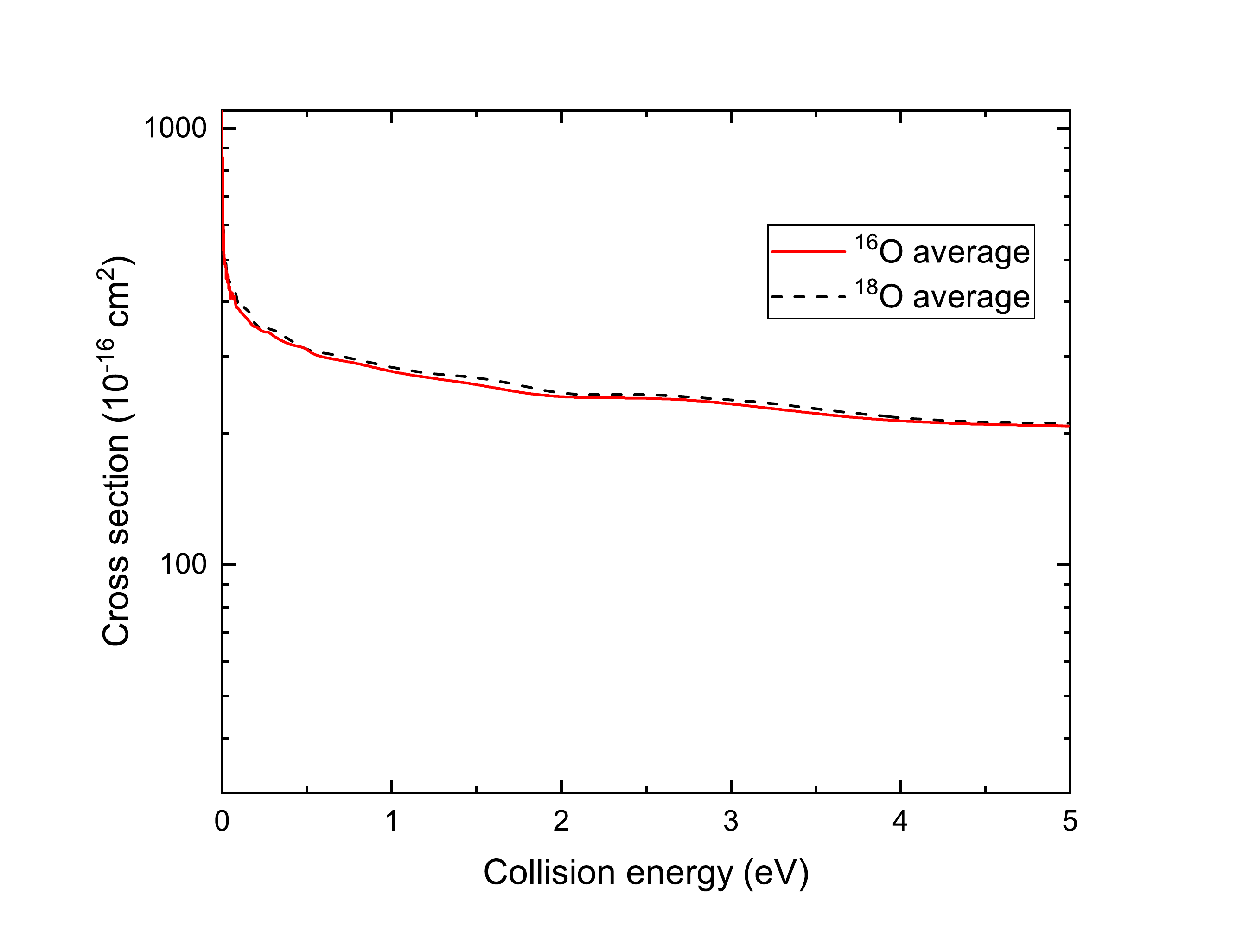}}
{\caption{\label{fig:elastic_compare_}Elastic cross sections for $^{16}$O+CO and $^{18}$O+CO scattering in dependence on the collision energy.}} 
\end{center}
\end{figure*}

\begin{figure*}  
\begin{center}
\includegraphics [height=0.50\textwidth]{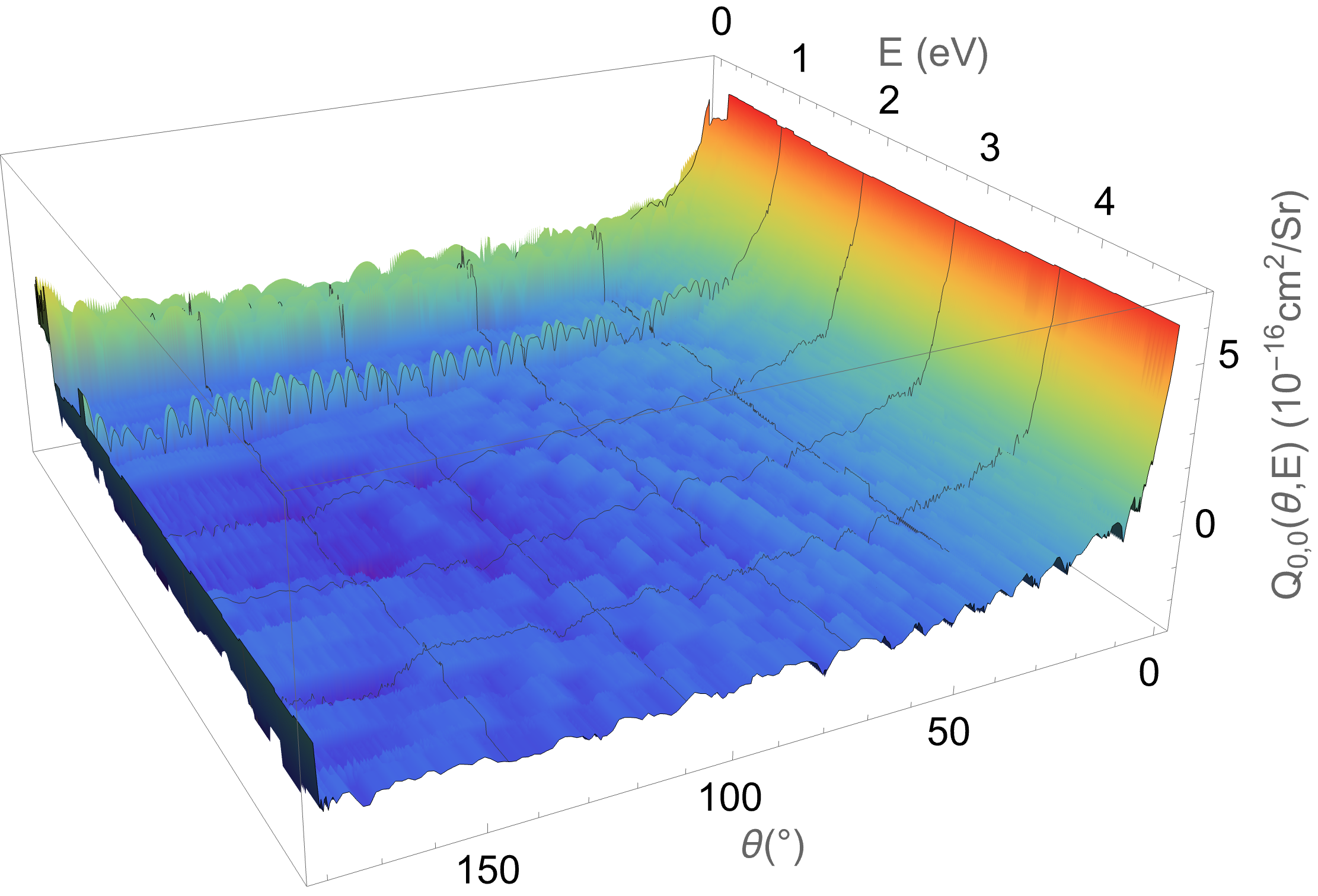}
{\caption{\label{fig:elastic_dcs_3d}Elastic differential cross section statistically averaged over all PES for $^{16}$O+CO scattering.}}
\end{center}
\end{figure*}

\begin{figure*}  
\begin{center}
\subfloat{\includegraphics [height=0.40\textwidth]{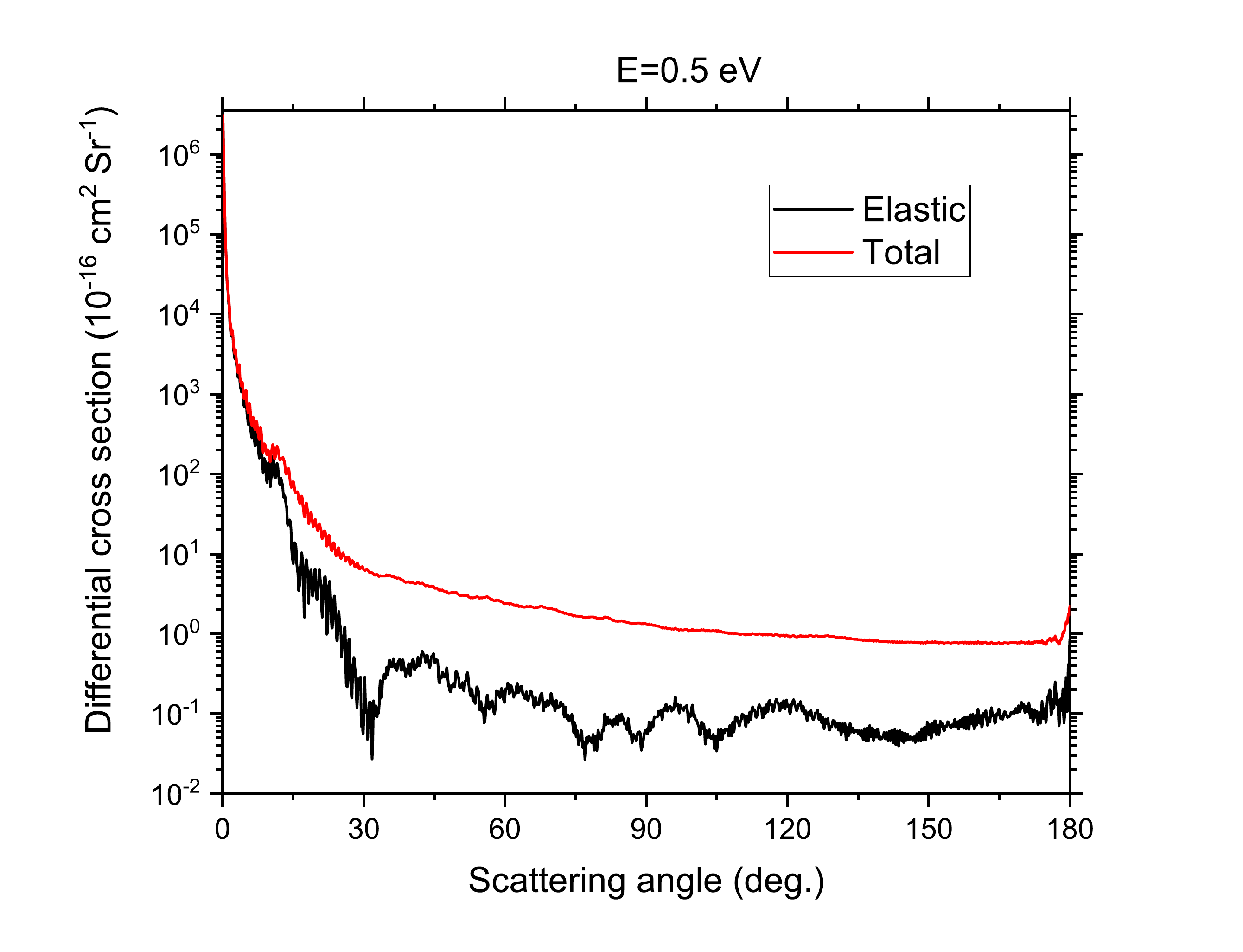}}
\subfloat{\includegraphics [height=0.40\textwidth]{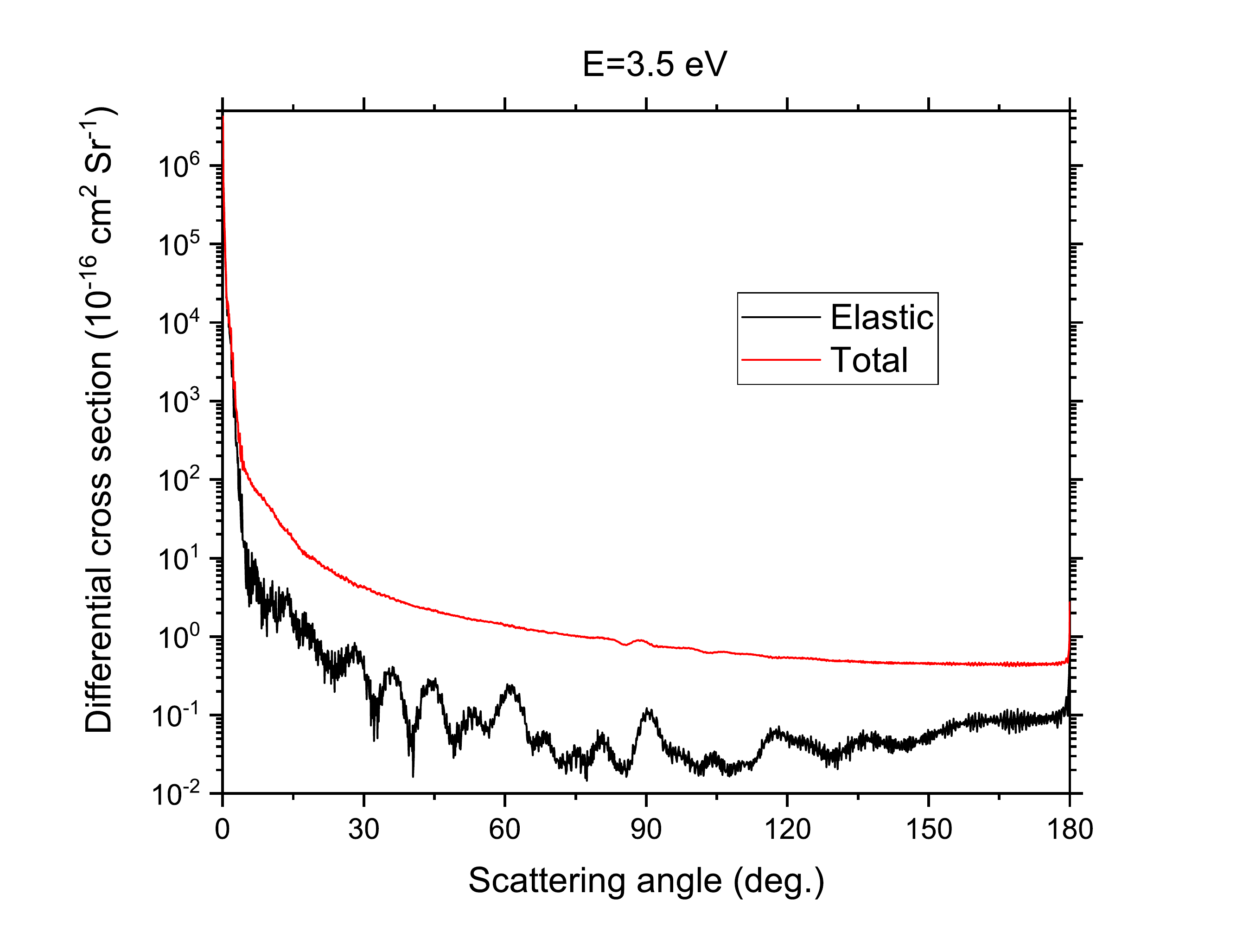}}
{\caption{\label{fig:total_dcs}Elastic and total differential cross sections for $^{16}$O+CO scattering at selected energies.}}
\end{center}
\end{figure*}

\begin{figure*} 
\begin{center}
\subfloat{\includegraphics [height=0.40\textwidth]{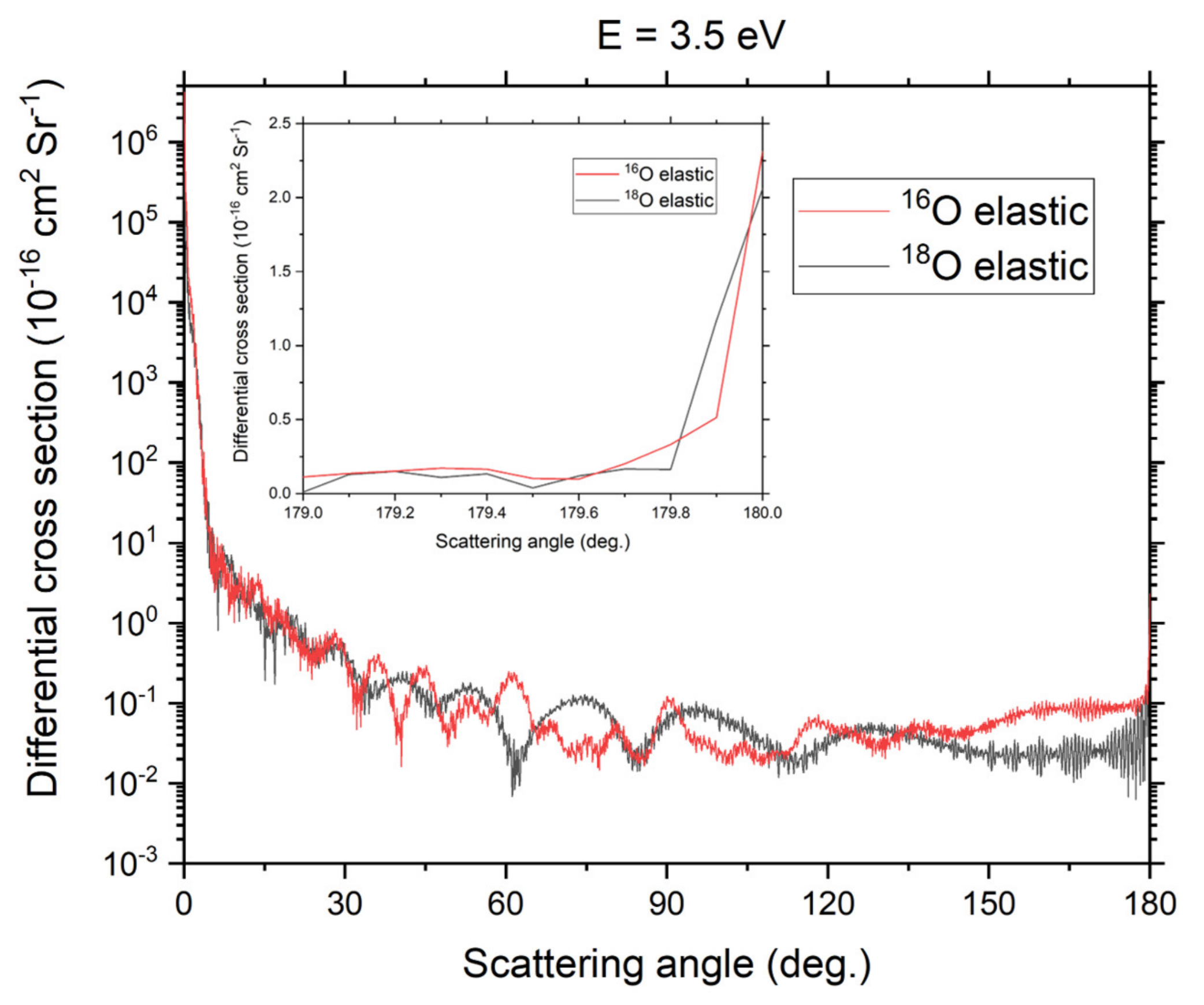}}
\subfloat{\includegraphics [height=0.40\textwidth]{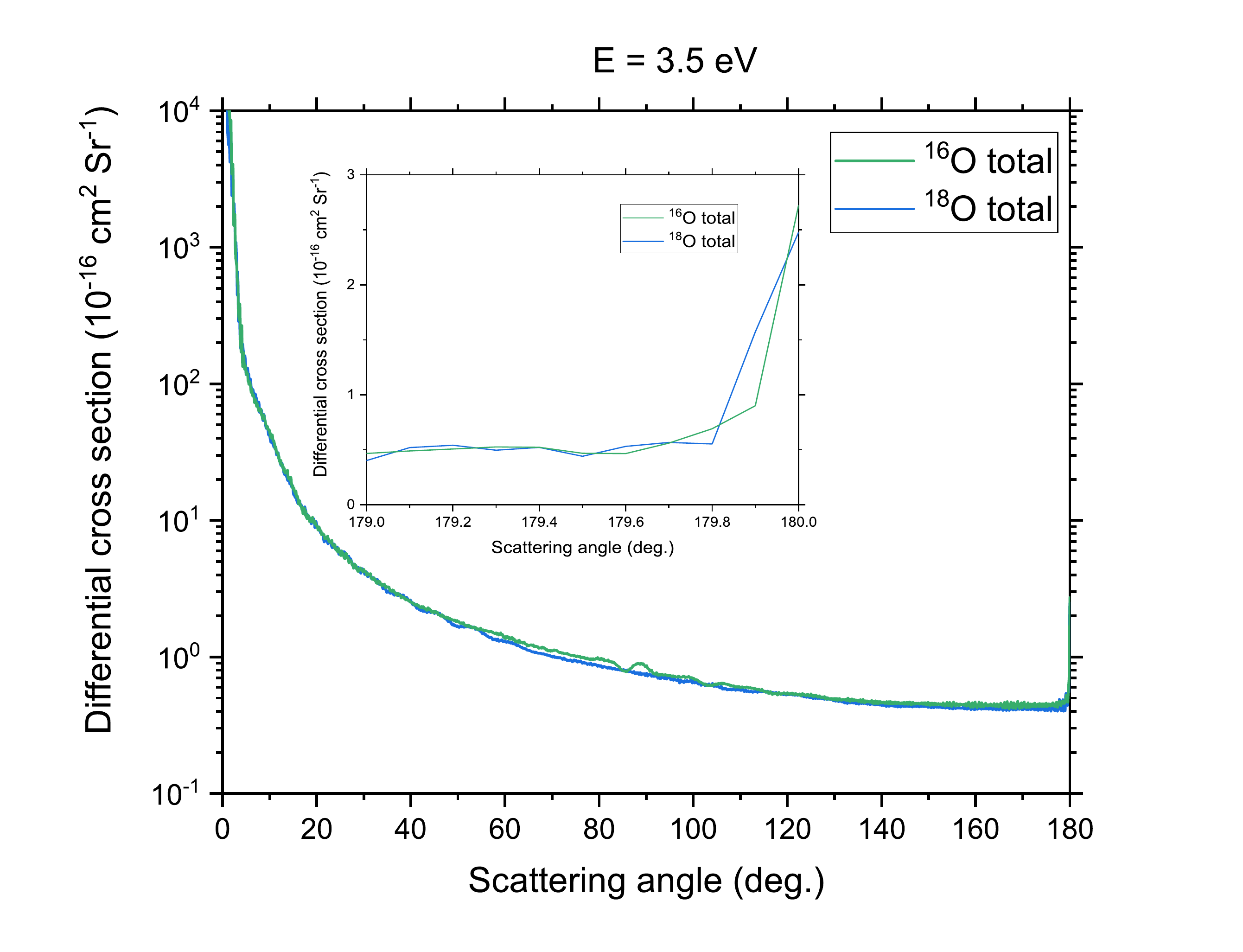}}
{\caption{\label{fig:comparison_16_18_3.5}Comparison of DCS of  $^{16}$O and $^{18}$O collisions with CO as a function of scattering angle at 3.5 eV}}
\end{center}
\end{figure*}


In order to facilitate a comparison with the values recommended and used in recent planetary aeronomy research \citep{lewkow2012energy,https://doi.org/10.1002/2016JA023461,FOX2018411,LO2021114371}, we performed a fit of the elastic cross sections in the following exponential form:
\begin{equation}{\label{eq06}}
   \sigma(E_k)=A E_k(\mathrm{eV})^{b} \, ,
\end{equation}
where $A$ and $b$ are the fitting parameters, and the collision energy $E_k$ is given in the units of eV. The fitting parameters obtained for the three considered isotopes of oxygen are given in Table \ref{tab:table2}.

\begin{table} 
\caption{\label{tab:table2}Fitting parameters for elastic cross sections as a function of energy (in eV) for $^{16}$O, $^{17}$O, and $^{18}$O with CO. Elastic cross sections for O+N$_2$ are given for comparison.}
\centering
\begin{tabular}{cccc}
\cline{1-4}
    Species       &   $A$ &  $b$ & Reference   \\ 
\hline
$^{16}$O-CO &    2.427 $\times$ 10$^{-14}$ &   -0.17196 & This work \\
$^{17}$O-CO &    2.761 $\times$ 10$^{-14}$ &   -0.16779 & This work \\
$^{18}$O-CO &    2.529 $\times$ 10$^{-14}$ &   -0.16557 & This work \\         
$^{16}$O-N$_2$ &  1.98 $\times$ 10$^{-14}$ &   -0.0816  & \cite{https://doi.org/10.1029/98JA02198}\\     
\cline{1-4}
\end{tabular}
\end{table}

The total cross sections as functions of collision energies were calculated using Eq. (\ref{eq03}) (Figure \ref{fig:elastic}). Specifically, a comparison of cross sections of $^{16}$O and  $^{18}$O collisions with CO is given in Figure \ref{fig:elastic_compare_}. These two cross sections exhibit similar energy dependence and the cross section for $^{18}$O-CO scattering remains larger than that for $^{16}$O-CO scattering throughout the entire collision energy range considered. This is an expected effect of the mass scaling between different isotopes due to the variations in the reduced mass.
\begin{table*} 
\caption{\label{tab:table5}State-to-state cross sections ($\sigma_{j=0,j'}$) (units of 10$^{-16}$ cm$^2$) for initial 30 rotational levels as a function of selected energies (in eV)  for $^{16}$O-CO scattering.}
\centering
\begin{tabular}{ccccccccccccc}
\cline{1-13}
  & & $E$ = 0.7 eV &  & & $E$ = 1.5 eV & & & $E$ = 3 eV & & & $E$ = 4.5 eV\\
$j'$ & $^3A'$ &  1 $^3A''$ & 2 $^3A''$ & $^3A'$ &  1 $^3A''$ & 2 $^3A''$ & $^3A'$ &  1 $^3A''$ & 2 $^3A''$ &  $^3A'$ &  1 $^3A''$ & 2 $^3A''$  \\
\cline{1-13} \\
 0     &     133.660    &666.370      &      78.553     &   117.557     &    596.057   &   62.914      &   122.759   &   525.377   &  52.453   &   120.937   &  460.320   &    47.546   \\
 1     &     3.238 &   18.789     &      1.569     &   5.794     &    18.996   &   0.951   &   7.034   &   14.966   &  0.725  &   6.683  &  11.941   &    0.54957                       \\
 2     &     7.316 &   11.281     &      6.741     &   5.666     &    8.337   &   5.743     &   5.620    &   5.459   &  5.048 &   5.063  &  4.677  &    5.023826                        \\
 3     &     6.467 &   6.361     &      0.603    &   7.308    &    3.395  &   0.638    &   7.116 &   2.248 &  0.718  &   6.060 &  2.043  &    0.616                                     \\
 4     &     3.359 &   3.786     &      4.148    &   1.342    &    2.428  &   3.611    &   1.221 &   1.953 &  3.053  &   1.229 &  2.015  &    3.053                                     \\
 5     &     4.233 &   3.809     &      0.697    &   3.932    &    2.921  &   0.812    &   2.595 &   2.495 &  0.764  &   2.318 &  2.346  &    0.748                                     \\
 6     &     3.497 &   1.649     &      2.921    &   1.231    &    1.199  &   2.423    &   0.726 &   0.885 &  2.079  &   0.727 &  0.768  &    1.889                                     \\
 7     &     1.530 &   2.407     &      0.705    &   1.790    &    2.248  &   0.985    &   1.036 &   1.991 &  1.086  &   0.957 &  1.776  &    0.997                                     \\
 8     &     2.129 &   1.188     &      2.133    &   0.989    &    0.807  &   1.790    &   1.075 &   0.646 &  1.447  &   1.078 &  0.540  &    1.366                                     \\
 9     &     1.222 &   1.473     &      1.186    &   1.041    &    1.589  &   1.368    &   0.819 &   1.501 &  1.348  &   0.661 &  1.471  &    1.164                                     \\
10     &     1.074 &   1.273     &      1.528    &   1.013    &    0.934  &   1.204    &   1.124 &   0.640 &  1.085  &   1.082 &  0.607  &    1.021                                     \\
11     &     1.545 &   0.950     &      1.227    &   1.039    &    0.970  &   1.247    &   0.490 &   1.005 &  1.337  &   0.466 &  0.992  &    1.223                                     \\
12     &     0.789 &   1.431     &      1.013    &   0.929    &    1.077  &   0.854    &   0.996 &   0.610 &  0.801  &   0.894 &  0.517  &    0.754                                     \\
13     &     1.031 &   0.620     &      1.388    &   0.603    &    0.691  &   1.366    &   0.405 &   0.730 &  1.313  &   0.521 &  0.678  &    1.309                                     \\
14     &     0.990 &   1.236     &      0.828    &   0.920    &    0.895  &   0.668    &   0.839 &   0.602 &  0.694  &   0.765 &  0.483  &    0.704                                     \\
15     &     0.627 &   0.430     &      1.378    &   0.394    &    0.497  &   1.365    &   0.398 &   0.569 &  1.176  &   0.499 &  0.615  &    1.248                                     \\
16     &     1.091 &   0.991     &      0.595    &   0.845    &    0.835  &   0.656    &   0.576 &   0.598 &  0.700  &   0.475 &  0.543  &    0.607                                     \\
17     &     0.335 &   0.435     &      1.284    &   0.392    &    0.498  &   1.234    &   0.341 &   0.576 &  1.036  &   0.424 &  0.484  &    1.056                                     \\
18     &     0.816 &   0.763     &      0.657    &   0.555    &    0.760  &   0.759    &   0.309 &   0.526 &  0.765  &   0.395 &  0.419  &    0.783                                     \\
19     &     0.161 &   0.482     &      1.098    &   0.382    &    0.371  &   0.954    &   0.570 &   0.450 &  0.899  &   0.463 &  0.493  &    0.853                                     \\
20     &     0.435 &   0.508     &      0.766    &   0.310    &    0.557  &   0.838    &   0.300 &   0.560 &  0.828  &   0.365 &  0.412  &    0.858                                     \\
21     &     0.258 &   0.550     &      0.832    &   0.582    &    0.333  &   0.755    &   0.557 &   0.441 &  0.793  &   0.384 &  0.373  &    0.707                                     \\
22     &     0.283 &   0.469     &      0.757    &   0.324    &    0.508  &   0.886    &   0.337 &   0.474 &  0.851  &   0.330 &  0.480  &    0.976                                     \\
23     &     0.346 &   0.569     &      0.536    &   0.678    &    0.275  &   0.598    &   0.367 &   0.497 &  0.717  &   0.255 &  0.411  &    0.498                                     \\
24     &     0.242 &   0.516     &      0.896    &   0.295    &    0.430  &   0.926    &   0.238 &   0.567 &  1.028  &   0.348 &  0.438  &    1.048                                     \\
25     &     0.300 &   0.580     &      0.331    &   0.478    &    0.331  &   0.521    &   0.290 &   0.312 &  0.583  &   0.325 &  0.427  &    0.554                                     \\
26     &     0.240 &   0.484     &      0.704    &   0.239    &    0.495  &   0.849    &   0.319 &   0.489 &  0.912  &   0.346 &  0.439  &    0.935                                     \\
27     &     0.228 &   0.571     &      0.239    &   0.276    &    0.370  &   0.514    &   0.274 &   0.518 &  0.685  &   0.334 &  0.400  &    0.776                                     \\
28     &     0.263 &   0.363     &      0.720    &   0.273    &    0.381  &   0.570    &   0.296 &   0.419 &  0.565  &   0.287 &  0.507  &    0.508                                     \\
29     &     0.189 &   0.310     &      0.108    &   0.257    &    0.342  &   0.417    &   0.202 &   0.361 &  0.558  &   0.217 &  0.495  &    0.588                                     \\
30     &     0.258 &   0.174     &      0.254    &   0.298    &    0.294  &   0.330    &   0.189 &   0.431 &  0.454  &   0.214 &  0.497  &    0.451                                     \\
\\
\cline{1-13}
\end{tabular}
\end{table*}

\begin{table} 
\caption{\label{tab:table3}Elastic cross sections ($\nu$=0, $j$=0) (units of 10$^{-16}$ cm$^2$) for the three PESs as a function of selected energies (in eV) for $^{16}$O+CO collisions.}
\centering
\begin{tabular}{ccccc}
\cline{1-5}
\\
$E$ (eV)       &   $^{3}A'$ &  1 $^{3}A''$ & $2^{3}A''$ &  Average  \\ \cline{1-5} \\
0.4&   119.321&  742.193& 91.754&  317.756 \\  
0.6&    127.700&  693.674 & 82.022&  301.132 \\ 
0.8&    135.169&  664.387 & 75.578&  291.711\\  
1.0&    126.358& 627.943 & 70.885&  275.062\\ 
1.5&    117.557&  596.057 & 62.914&  258.842 \\ 
2.0&    118.031&  548.743 & 58.206&  241.660 \\ 
2.5&    120.209&  547.809 & 54.933&  240.983 \\ 
3.0&    122.759&  525.377 & 52.453&  233.529 \\ 
3.5&    124.312& 491.953 & 50.154&  222.139 \\ 
4.0&    122.945&  469.177 & 48.977&213.699 \\ 
4.5&    120.937&  460.320 & 47.546&  209.601\\ 
5.0&    117.767& 458.113 & 45.001&  206.960 \\ 
\cline{1-5}
\end{tabular}
\end{table}

\subsection{Differential cross sections}

We calculated elastic and total (elastic+inelastic) differential cross sections (DCSs) from the scattering matrices using the utility code \textsl{dcssave.f} included in the MOLSCAT program package \citep{hutson1994molscat}. The final DCSs for the three isotopes of O considered were calculated from the statistically averaged sum of the individual molecular channels.
Figure \ref{fig:elastic_dcs_3d} shows the differential cross section for $^{16}$O+CO elastic collisions as a function of the collision energy and scattering angle. Cuts through the DCS surface for specific energies are given in Figure \ref{fig:total_dcs} for 0.5 eV and 3.5 eV, respectively. 

The role of isotope fractionation in atmospheric escape is important for understanding the evolution of planetary atmospheres. The isotope dependence of the DCSs was investigated and the results are shown in Figure \ref{fig:comparison_16_18_3.5}. 
Specifically, Figure \ref{fig:comparison_16_18_3.5} shows DCSs for $^{16}$O+CO and $^{18}$O+CO collisions as a function of scattering angle at 3.5 eV. Both elastic and total DCSs are highly anisotropic, with predominant forward-scattering peak and noticeable backscattering features for large scattering angles. The backscattering for $\theta=179^{\circ}$ to $\theta=180^{\circ}$ is illustrated in Figure \ref{fig:comparison_16_18_3.5} (inset). Note that the integral cross sections were computed for all the scattering angles in the range of 0$^{\circ}$ $-$ 180$^{\circ}$ with an angular resolution of 0.1$^{\circ}$. 
In all cases, the differential cross sections are nearly elastic at low scattering angles, up to about $\theta=7^{\circ}$. For larger scattering angles, the total DCSs are up to two orders of magnitude larger than the elastic cross sections indicating the dominant role of rotational excitations, especially at high collision energies. The fast oscillations present in the computed DCSs are due to the quantum-mechanical interferences. 
Note that while here we present the data for the $^{16}$O-CO and $^{18}$O-CO collisions, we have calculated the elastic and total cross sections and the associated differential cross sections for the three isotopes of oxygen considered, namely $^{16}$O-CO, $^{17}$O-CO, and $^{18}$O-CO. The complete scattering datasets for $^{17}$O-CO, along with $^{16}$O and $^{18}$O, is given in a digital repository made available online \citep{github_repo,sanchit_chhabra_2022_7084641}.

\section{SUMMARY AND DISCUSSION}
We carried out a quantum dynamics study of $^{16}$O($^3P$)+CO, $^{17}$O($^3P$)+CO, and $^{18}$O($^3P$)+CO collisions at energies up to 5 eV, as defined in the center-of-mass frame. The O($^3P$)-CO electronic interactions were described using the three lowest-energy electronic potential energy surfaces correlating to the first asymptote of the interacting OCO complex by \citet{schwenke2016collisional}. For the three isotopes of oxygen, the velocity-dependent state-to-state and total elastic and inelastic integral cross sections as well as corresponding differential cross sections were calculated. The CO molecule was modeled in the rigid rotor approximation (Arthurs-Dalgarno model) and the collisions were described as non-reactive. In order to make the computational problem tractable, the coupled state (CS) approximation \citep{doi:10.1063/1.1681388} was invoked. Thus, the cross sections and derived quantities were constructed from the first principles with no fitting parameters.

The rigid rotor approximation is justified in the physical regime found in the irradiated upper planetary atmospheres of CO$_2$-rich worlds and similar non-thermal environments where the attenuation of the superthermal oxygen atoms in CO collisions occurs. This is evidenced by the strongly forward-peaked differential cross sections which suggest that the inelastic collisions with significant energy transfer required for vibrational excitations can occur only at large scattering angles. This is expected behavior in atom-atom and atom-molecule collisions at superthermal collision energies studied in the context of fast atom thermalization \citep{1982itam.book.....J,KHARCHENKO1997107,https://doi.org/10.1029/2000JA000085,10.1093/mnras/stz3366}. Similar to our conclusions, \citet{brunsvold_upadhyaya_zhang_cooper_minton_braunstein_duff_2008} reported a mechanism for large energy transfers in inelastic O-CO collisions, into both rotations and vibrations of the target CO molecule, that takes place for large scattering angles, $\theta=80^\circ$ to $\theta=140^\circ$, that correspond to a bending mode of OCO complex. Additionally, \citet{brunsvold_upadhyaya_zhang_cooper_minton_braunstein_duff_2008} estimated that in collisions at 80 kcal/mol (3.47 eV) about 15\% of the collision energy is transferred to the excitations of internal degrees of CO.  
Overall, our computed inelastic DCSs are somewhat larger than reported by \citet{brunsvold_upadhyaya_zhang_cooper_minton_braunstein_duff_2008}, which is likely due to our choice of the computational approach and the differences in the potential energy surfaces. \cite{doi:10.1063/1.462105} determined the excitation cross sections of fast O-atom collisions with CO. The resulting cross section for O+CO is found to be 7.3  $\times$ 10$^{-17}$ cm$^2$, for a center of mass energy of 3.4 eV. In the absence of cross sections data, the mass scaling procedures are used to determine the unknown cross sections for atom–atom and atom–molecule collisions \citep{FOX2018411}. 

The investigated collision energy range was selected primarily to address the collisions with non-thermal O atoms and kinematics at non-thermal equilibrium conditions in upper planetary atmospheres of CO$_2$-rich planets, such as Mars and Venus as well as exoplanets, such as WASP-39b where the presence of CO$_2$ was recently confirmed \citep{early2022identification}. 
At Mars, the photochemical escape of oxygen to space is a major loss mechanism in the present epoch \citep{https://doi.org/10.1002/2016JA023461} and a potential climate evolution driver, as well as a mediating process capable of strongly affecting the escape of atomic carbon \citep{LO2021114371}. Similarly to O($^3P$)+CO$_2$ \citep{10.1093/mnras/stz3366}, the O($^3P$)+CO cross sections are responsible for determining attenuation of the neutral oxygen escape at higher altitudes, where CO is more abundant \citep{https://doi.org/10.1029/2001JE001809}. Similarly, at Venus, the O+CO cross sections are one of the important parameters in the models of the non-thermal hot oxygen environment \citep{https://doi.org/10.1029/2010JE003697}.

\section*{Acknowledgements}
SC and NEK were partly supported by Khalifa University (grants CIRA-2019-054, AARE-000329-00001 and 8474000336-KU-SPSC). MG acknowledges support from Khalifa University (grants 8474000362-KU-FSU-2021 and 8474000336-KU-SPSC).

%



\bibliographystyle{mnras}
\bibliography{OCO_bibliography}




%
%


\bsp	
\label{lastpage}
\end{document}